\documentstyle[12pt]{article}
\newcommand{\RR}{$\rm l\!R\;\;$}

\newcommand{\eut}{
\begin{picture}(3,3)(-2,-2)
\put(-2,-15){$\tilde{}$}
\put(-5,-2){$\eta$}
\end{picture}}

\begin{document}
\hspace{10.8cm} CGPG-94/10-2
\begin{center}

\baselineskip=24pt plus 0.2pt minus 0.2pt
\lineskip=22pt plus 0.2pt minus 0.2pt

 \Large

Reality Conditions and Ashtekar Variables: a Different Perspective

\vspace*{0.35in}

\large

J.\ Fernando\ Barbero\ G. $^{\ast, \dag}$
\vspace*{0.25in}

\normalsize

$^{\ast}$Center for Gravitational Physics and Geometry\\
Department of Physics,\\
Pennsylvania State University,\\
University Park, PA 16802\\
U.S.A.\\

$^{\dag}$Instituto de Matem\'aticas y F\'{\i}sica Fundamental, \\
C.S.I.C.\\
Serrano 119--123, 28006 Madrid, Spain
\\

\vspace{.5in}
October 10, 1994\\
\vspace{.5in}

ABSTRACT

\end{center}
\vspace{.5in}

We give in this paper a modified self-dual action that leads to the
$SO(3)$-ADM formalism without having to face the difficult second class
constraints present in other approaches (for example, if one starts from
the Hilbert-Palatini action). We use the new action principle to gain some
new insights into the problem of the reality conditions that must be
imposed in order to get real formulations from complex general relativity.
We derive also a real formulation for Lorentzian general relativity in the
Ashtekar phase space by using the modified action presented in the paper.


\pagebreak

\baselineskip=24pt plus 0.2pt minus 0.2pt
\lineskip=22pt plus 0.2pt minus 0.2pt

\setcounter{page}{1}

\section{Introduction}

The purpose of this paper is to present a modified form of the self-dual
action and use it to discuss the problem of reality conditions in the
Ashtekar description of general relativity. By now, the Ashtekar
formulation \cite{alfa} has provided us with a new way to study gravity
from a non-perturbative point of view. The success of the program can be
judged from the literature available about it \cite{beta}. In our opinion
there are two main technical points that have contributed to this success.
The first one is the fact that the configuration variable is an $SO(3)$
connection. This allows us to formulate general relativity in the familiar
phase space of the Yang-Mills theory for this group. We can then take
advantage of the many results about connections available in the
mathematical physics literature. In particular, it proves to be very
useful to have the possibility of using loop variables \cite{beeta}
(essentially Wilson loops of the Ashtekar connection and related objects)
in both the classical and the quantum descriptions of the theory. A second
important feature of the Ashtekar formalism is the fact that the
constraints (in particular the Hamiltonian constraint) have a very simple
structure when written in terms of the new variables.  This has been very
helpful in order to find solutions to all the constraints of the theory
and is in marked contrast with the situation in the ADM formalism
\cite{gamma} where the scalar constraint is very difficult to work with
because of its rather complicated structure.

In spite of all the success of the formulation, there are still several
problems that the Ashtekar program has to face. The one that we will be
mostly concerned with in this paper is the issue of the reality
conditions. As it is well known, the so called reality conditions must be
imposed on the complex Ashtekar variables in order to recover the usual
real formulation of general relativity for space-times with Lorentzian
signatures. Their role is to guarantee that both the three-dimensional
metric and its time derivative (evolution under the action of the
Hamiltonian constraint) are real. This introduces key difficulties in the
formulation, specially when one tries to work with loop variables
(although some progress on this issue has been recently reported
\cite{delta}).

The main purpose of this paper is to clarify some issues related with the
real formulations of general relativity that can be obtained from a given
complex theory. We will see, for example, that both in the\footnote {in
the following we mean by $SO(3)$-ADM formalism the version of the ADM
formalism in which an internal $SO(3)$ symmetry group has been introduced
as in \cite{eta}.} $SO(3)$-ADM and in the Ashtekar phase space it is
possible to find Hamiltonian constraints that trivialize the reality
conditions to be imposed on the complex theory (regardless of the
signature of the space-time). Conversely, any of this alternative forms
for the constraints in a given phase space can be used to describe
Euclidean or Lorentzian space-times, provided that we impose suitable
reality conditions. Though this fact is, somehow, obvious in the ADM
framework, it is not so in the Ashtekar formalism. In doing this we will
find a real formulation for Lorentzian general relativity in the Ashtekar
phase space. The main difference between this formulation and the more
familiar one is the form of the scalar constraint. We will need a
complicated expression in order to describe Lorentzian signature
space-times.  In our approach, the problem of the reality conditions is,
in fact, transformed into the problem of writing the new Hamiltonian
constraint in terms of loop variables and, in the Dirac quantization
scheme, imposing its quantum version on the wave functionals (issues that
will not be addressed in this paper). Of course one must also face the
difficult problems of finding a scalar product in the space of physical
states etc...

A rather convenient way of obtaining the new Hamiltonian constraint is by
starting with a modified version of the usual self-dual action
\cite{epsilon} that leads to the $SO(3)$-ADM formalism in such a way that
the transition to the Ashtekar formulation is very transparent. We will
take advantage of this fact in order to obtain the real Lorentzian
formulation and to discuss the issue of reality conditions.

The lay-out of the paper is as follows. After this introduction we review,
in section II, the self-dual action and rewrite it as the
Husain-Kucha\v{r} \cite{zeta} action coupled to an additional field. This
will be useful in the rest of the paper. Section III will be devoted to
the modified self-dual action that leads to the $SO(3)$-ADM formalism. We
discuss the issue of reality conditions in section IV. We will show that
although multiplying the usual self-dual action by a purely imaginary
constant factor does not change anything (both at the level of the field
equations and the Hamiltonian formulation), the same procedure, when used
with the modified self-dual action changes the form of the ADM Hamiltonian
constraint (in fact it changes the relative sign between the kinetic and
potential terms that in a real formulation controls the signature of the
space-time). In section V we derive the real Ashtekar formulation for
Lorentzian signatures and we end the paper with our conclusions and
comments in section VI.

\section{The self dual action and Ashtekar variables}

We will start by introducing our conventions and notation. Tangent space
indices and $SO(3)$ indices are represented by lowercase Latin letters
from the beginning and the middle of the alphabet respectively. No
distinction will be made between 3-dimensional and 4-dimensional tangent
space indices (the relevant dimensionality will be clear from the
context). Internal $SO(4)$ indices are represented by capital latin
letters from the middle of the alphabet. The 3-dimensional and
4-dimensional Levi-Civita tensor densities will be denoted\footnote{We
represent the density weights by the usual convention of using tildes
above and below the fields.} by $\tilde{\eta}^{abc}$ and
$\tilde{\eta}^{abcd}$ and the internal Levi-Civita tensors for both
$SO(3)$ and $SO(4)$ represented by $\epsilon_{ijk}$ and $\epsilon_{IJKL}$.
The tetrads $e_{aI}$ will be written in components as $e_{aI}\equiv
(v_{a}, e_{ai})$ (although at this point the i index only serves the
purpose of denoting the last three internal indices of the tetrad we will
show later that it can be taken as an $SO(3)$ index). $SO(4)$ and $SO(3)$
connections will be denoted by $A_{aIJ}$ and $A_{ai}$ respectively with
corresponding curvatures $F_{abIJ}$ and $F_{abi}$ given by $F_{abIJ}\equiv
2\partial_{[a} A_{b]IJ}+A_{aI}^{\;\;\;\;K}A_{bKJ}-A_{bI}^{\;\;\;\;K}
A_{aKJ}$ and $F_{ab}^{i}\equiv 2\partial_{[a}
A_{b]}^{i}+\epsilon^{i}_{\;\;jk}A_{a}^{j} A_{b}^{k}$. The actions of the
covariant derivatives defined by these connections on internal indices are
$\nabla_{a}\lambda_{I}=\partial_{a}\lambda_{I}+A_{aI}^{\;\;\;\;K}\lambda_{K}$
and $\nabla_{a}\lambda_{i}=\partial_{a}\lambda_{i}+\epsilon_{ijk}A_{aj}
\lambda_{k}$. They can be extended to act on tangent space indices by
introducing a torsion-free connection (for example the Christoffel
connection $\Gamma_{ab}^{c}$ built with the four-metric $q_{ab}\equiv
e_{aI}e^{I}_{b}$). All the results in the paper will be independent of
such an extension. We will work with self-dual and anti-self-dual objects
satisfying $B^{\pm}_{IJ}=\pm\frac{1}{2}
\epsilon_{IJ}^{\;\;\;\;KL}B^{\pm}_{KL}$ where we raise and lower $SO(4)$
indices with the internal Euclidean metric Diag(++++). In particular,
$A^{-}_{IJ}$ will be an anti-self-dual $SO(4)$ connection (taking values
in the anti-selfdual part of the complexified Lie algebra of $SO(4)$) and
$F_{abIJ}^{-}$ its curvature. In space-times with Lorentzian signature a
factor {\it i} must be included in the definition of self-duality if we
impose the usual requirement that the duality operation be such that its
square is the identity and raise and lower internal indices with the
Minkowski metric Diag($-$+++). In this paper we will consider complex
actions invariant under complexified $SO(4)$. For the purpose of
performing the 3+1 decomposition the space-time manifold is restricted to
have the form ${\cal M}= $\RR$ \!\!\!\times\Sigma$ with $\Sigma$ a compact
3-manifold with no boundary.

The Samuel-Jacobson-Smolin \cite{epsilon} action is
\begin{equation}
S=\int_{\cal M} d^{4}\!x \;\tilde{\eta}^{abcd}F^{-IJ}_{ab} e_{cI}e_{dJ}
\label{1}\\
\end{equation}
It is useful to rewrite it in a
slightly modified manner \cite{theta}. We start by writing the
anti-self-dual connection and the tetrad in matrix form as
\begin{equation} \begin{array}{cc} A^{-}_{aIJ}\equiv\frac{1}{2}\left[
\begin{array}{rrrr}
                              0 & A_{a}^{1} & A_{a}^{2} & A_{a}^{3} \\
                              -A_{a}^{1} & 0 & -A_{a}^{3} & A_{a}^{2} \\
                              -A_{a}^{2} & A_{a}^{3} & 0 & -A_{a}^{1} \\
                              -A_{a}^{3} & -A_{a}^{2} & A_{a}^{1} & 0
                              \end{array}
                      \right] &
\hspace{1cm} e_{a}^{I}\equiv\left[  \begin{array}{c}
                             v_{a} \\ e_{a}^{1} \\e_{a}^{2} \\ e_{a}^{3}
                  \end{array} \right]
\end{array}
\label{2}
\end{equation}
Under anti-self-dual and self-dual $SO(4)$ infinitesimal transformations
generated by
\begin{equation}
\begin{array}{cc}
\Lambda^{-}_{IJ}=\left[  \begin{array}{rrrr}
                    0 & \Lambda_{a}^{1} & \Lambda_{a}^{2} & \Lambda_{a}^{3} \\
                  -\Lambda_{a}^{1} & 0 & -\Lambda_{a}^{3} & \Lambda_{a}^{2} \\
                  -\Lambda_{a}^{2} & \Lambda_{a}^{3} & 0 & -\Lambda_{a}^{1} \\
                  -\Lambda_{a}^{3} & -\Lambda_{a}^{2} & \Lambda_{a}^{1} & 0
                              \end{array}
                      \right] &
\Lambda^{+}_{IJ}=\left[  \begin{array}{rrrr}
                              0 & L_{a}^{1} & L_{a}^{2} & L_{a}^{3} \\
                              -L_{a}^{1} & 0 & L_{a}^{3} & -L_{a}^{2} \\
                              -L_{a}^{2} & -L_{a}^{3} & 0 & L_{a}^{1} \\
                              -L_{a}^{3} & L_{a}^{2} & -L_{a}^{1} & 0
                              \end{array}
                      \right]
\end{array}
\label{3}
\end{equation}
the fields transform as
\begin{equation}
\begin{array}{l}
\begin{array}{l}
\delta^{-}(\Lambda)\;A^{-}_{aIJ}=-\partial_{a}\Lambda_{IJ}-A_{aI}^{\;\;\;\;K}
\Lambda_{KJ}+A_{aJ}^{\;\;\;\;K}\Lambda_{KI} \\
\delta^{-}(\Lambda)\;v_{a}=\Lambda_{i}e_{a}^{i} \\
\delta^{-}(\Lambda)\;e_{ai}=-\Lambda_{i}v_{a}-\epsilon_{ijk}e_{a}^{j}
\Lambda^{k}
\end{array}
                             \\ \\
\begin{array}{l}
\delta^{+}(L)\;A^{-}_{aIJ}=0 \\
\delta^{+}(L)\;v_{a}=L_{i}e_{a}^{i} \\
\delta^{+}(L)\;e_{ai}=-L_{i}v_{a}+\epsilon_{ijk}e_{a}^{j}L^{k}
\end{array}

\end{array}
\label{4}
\end{equation}
The transformations of the connections can be written also as
\begin{equation}
\begin{array}{ll}
\delta^{-}(\Lambda)\;A_{ai}=-2(\partial_{a}\Lambda_{i}+\epsilon_{i}^{\;\;jk}
A_{aj}\Lambda_{k}) & \hspace{1cm} \delta^{+}(L)\;A_{ai}=0
\end{array}
\label{5}
\end{equation}
It is easy to show that $\delta^{1}$ and
$\delta^{+}$ are two sets of commuting $SO(3)$ transformations
corresponding to the
factors in $SO(4)=SO(3)\bigotimes SO(3)$.
The transformation law of $A_{ai}$ under anti-self-dual $SO(4)$
transformations is that of an $SO(3)$ connection\footnote{This is the
reason why we introduced anti-self-dual connections in the action
(\ref{1}).} but that of the rest of the fields is not (i.e. we can not
take $i, j, k...$ as $SO(3)$ indices at this stage). However, by
considering simple combinations of self-dual and anti-self-dual
transformations $\delta^{1}(M)\equiv \delta^{-}(M/2)-\delta^{+}(M/2)$, we
have

\begin{equation}
\begin{array}{l}
\delta^{1}(M)\;A_{ai}=-(\partial_{a}M_{i}+\epsilon_{i}^{\;\;jk}A_{aj}M_{k})\\
\delta^{1}(M)\;v_{a}=0\\
\delta^{1}(M)\;e_{ai}=-\epsilon_{ijk}e_{a}^{j}M^{k}
\end{array}
\label{6}
\end{equation}

As we can see, $A_{ai}$, $e_{ai}$, $v_{a}$ do transform as $SO(3)$ objects
under the action of $\delta^{1}$ if we consider the indices $i, j, k...$
as $SO(3)$ indices.  The invariance of $v_{a}$ under these transformations
makes it very natural to consider the gauge fixing condition $v_{a}=0$
that we will use later.
In terms of $A_{ai}$, $v_{a}$ and $e_{ai}$ the action
(\ref{1}) reads
\begin{equation}
S=\int_{{\cal M}} d^{4}\!x \tilde{\eta}^{abcd}\left[
v_{a}e_{bi}F_{cdi}-\frac{1}{2}\epsilon^{ijk}e_{ai}e_{bj}F_{cdk}\right]
\label{7}
\end{equation}

This form of the Samuel-Jacobson-Smolin action has some nice features. It
shows, for example, that general relativity can be obtained from the
Husain-Kucha\v{r} \cite{zeta} model action by introducing a vector field
$v_{a}$ and a suitable interaction term. This is useful in order to study
the dynamics of degenerate solutions given by the action (in contrast with
the usual approach of extending the validity of the Ashtekar constraints
to the degenerate sector of the theory). The action (\ref{7}) will also be
the starting point of the next section in which we show that a certain
modification of it gives rise to the $SO(3)-$ADM formalism and provides a
very natural way of linking it to the Ashtekar formulation.

The fact that complexified $SO(4)$ and $SO(1,3)$ coincide means that we
can start from (\ref{1}), raise and lower indices with the Minkowski
metric Diag($-$+++) and define self-dual and anti-self-dual fields by
$B^{\pm}_{IJ}=\pm\frac{i}{2} \epsilon_{IJ}^{\;\;\;\;KL}B^{\pm}_{KL}$. It
is straightforward to show that the resulting action is equivalent to
(\ref{1}) because they can be related by simple redefinitions of the
fields.

In the passage to the Hamiltonian formulation \footnote{We include this
short discussion for further reference; the details can be found in
\cite{epsilon}.} corresponding to (\ref{7}) we introduce a foliation of
the space-time manifold ${\cal M}$ defined by hypersurfaces of constant
value of a scalar function t. We need also a congruence of curves with
tangent vector $t^{a}$ satisfying $t^{a}\partial_{a}t=1$ (with this last
requirement time derivatives can be interpreted as Lie derivatives ${\cal
L}_{t}$ along the vector field $t^{a}$). Performing the 3+1 decomposition
we have
$$
S={\displaystyle \int} dt{\displaystyle \int}_{\Sigma} d^{3}\!x \left\{
({\cal L}_{t} A_{a}^{i}) \tilde{\eta}^{abc}\left[
2v_{b}e_{ci}-\epsilon_{i}^{\;\;jk}
e_{bj}e_{ck}\right]+A_{0}^{i}\nabla_{a}\left[\tilde{\eta}^{abc}(2v_{b}e_{ci}-
\epsilon_{i}^{\;\;jk}e_{bj}e_{ck})\right]+\right.
$$
\begin{equation}
\left.\hspace{-3.3cm} +v_{0}\tilde{\eta}^{abc}e_{a}^{i}F_{bci}-e_{0}^{i}
\tilde{\eta}^{abc}
\left[v_{a}F_{bci}+\epsilon_{i}^{\;\;jk}e_{aj}F_{bck}\right]\right\}\equiv
{\displaystyle \int} dt {\cal L}(t)
\label{8}
\end{equation}
where $A_{0}^{i}\equiv t^{a}A_{a}^{i}$, $e_{0}^{i}\equiv t^{a}e_{a}^{i}$,
and $v_{0}\equiv t^{a}v_{a}$. All the objects in (\ref{8}) are effectively
three-dimensional (they can be taken as tensors in the spatial
hypersurfaces $\Sigma$). Denoting as $\tilde{\pi}^{a}_{i}(x)$,
$\tilde{\pi}_{i}(x)$, $\tilde{\sigma}^{a}_{i}(x)$,
$\tilde{\sigma}_{i}(x)$, $\tilde{p}^{a}(x)$, and $\tilde{p}(x)$ the
momenta canonically conjugate to $A_{a}^{i}(x)$, $A_{0}^{i}(x)$,
$e_{a}^{i}(x)$, $e_{0}^{i}(x)$, $v_{a}(x)$, and $v_{0}(x)$ respectively
($\{A_{a}^{i}(x),
\pi^{b}_{j}(y)\}=\delta_{a}^{b}\delta_{j}^{i}\delta^3(x,y)$, and so on) we
get from (\ref{8}) the following primary constraints
\begin{equation}
\hspace{-4.7cm}
\begin{array}{l}
\tilde{\pi}_{i}=0 \\
\tilde{\sigma}_{i}=0 \\
\tilde{p}=0
\end{array}
\label{10}
\end{equation}
\begin{equation}
\begin{array}{l}
\tilde{\pi}_{i}^{a}-\tilde{\eta}^{abc}(2v_{b}e_{ci}-\epsilon_{i}^{\;\;jk}
e_{bj}e_{ck})=0 \\
\tilde{\sigma}_{i}^{a}=0 \\
\tilde{p}^{a}=0
\end{array}
\label{10b}
\end{equation}
The constraints (\ref{10}) are first class, whereas (\ref{10b}) are second
class. The conservation in time of these constraints gives the secondary
constraints
\begin{eqnarray}
& & \nabla_{a}\left[\tilde{\eta}^{abc}(2v_{b}e_{ci}-
\epsilon_{i}^{\;\;jk}e_{bj}e_{ck})\right]=0\nonumber\\
& & \tilde{\eta}^{abc}
\left[v_{a}F_{bci}+\epsilon_{i}^{\;\;jk}e_{aj}F_{bck}\right]=0\label{13}\\
& & \tilde{\eta}^{abc}e_{a}^{i}F_{bci}=0\nonumber
\end{eqnarray}
that added to the set of primary constraints are second class. It is
possible to show, at least when the triads are non-degenerate, that
$v_{a}$ is pure gauge and so we can consistently remove both $v_{a}$ and
$\tilde{p}^{a}$ from all the expressions of the constraints (see
\cite{iota} for details on this issue). The price that we pay is that we
will not find the generator of the full $SO(4)$ in the final Hamiltonian
formulation but only one of the $SO(3)$ factors. From here, following the
usual steps of Dirac's \cite{Dir} procedure to deal with constrained
systems one gets the familiar Ashtekar constraints
\begin{eqnarray}
\nabla_{a}\tilde{\pi}^{a}_{i}=0\nonumber\\
\tilde{\pi}^{a}_{i}F_{ab}^{i}=0\label{0014}\\
\epsilon^{ijk}\tilde{\pi}^{a}_{i}\tilde{\pi}^{b}_{j}F_{abk}=0
\nonumber
\end{eqnarray}
where $A_{a}^{i}$ and $\tilde{\pi}^{a}_{i}$ are a canonically conjugate
pair of variables.

\section{The modified self-dual action}

We show in this section that a simple modification of the action (\ref{7})
gives a theory with Hamiltonian formulation given by the $SO(3)$-ADM
formalism (see \cite{wal} for a proposal somehow related to ours). The
derivation of this result is easier than in the case of starting from the
Palatini action as in \cite{eta} because the second class constraints are
much simpler to deal with. This result is interesting for several reasons.
It will be used in the next section to discuss the reality conditions of
the theory. It leads also in a very natural way to some of the real
Hamiltonian formulations for Lorentzian general relativity in the Ashtekar
phase space discussed in (\cite{fer1}). Throughout this section all the
fields will be taken as complex.

The key idea to get the modified action is realizing that
$\tilde{\eta}^{abcd}\epsilon^{ijk}e_{ai}e_{bj}F_{cdk}=-2\tilde{\eta}^{abcd}
e_{ai}\nabla_{b}\nabla_{c}e_{di}=-2\tilde{\eta}^{abcd}\left[\nabla_{b}
(e_{a}^{i}\nabla_{c}e_{di})+(\nabla_{a}e_{bi})(\nabla_{c}e_{di})\right]$.
By adding, then, a total derivative to (\ref{7}) we get

\begin{equation}
S=\int_{{\cal M}}
d^{4}\!x\;\tilde{\eta}^{abcd}\left[(\nabla_{a}e_{bi})(\nabla_{c}e_{di})+
v_{a}e_{b}^{i}F_{cdi}\right]
\label{2.1}
\end{equation}

In doing this we are, in fact, using the familiar procedure to
generate canonical transformations by adding a divergence to the
Lagrangian. The term introduced in order to get (\ref{2.1}) can be found
in \cite{Dol} and is given by

\begin{equation}
\int
d^3\!x\;\tilde{\eta}^{abc}\epsilon_{ijk}(A_{a}^{i}-\Gamma_{a}^{i})
e_{b}^{j}e_{c}^{k}=
\int
d^3\!x\;\tilde{\eta}^{abc}e_{a}^{i}\nabla_{b}e_{ci}
\label{0015}
\end{equation}
in this last expression we have used the compatibility of $\Gamma_{a}^{i}$
and $e_{a}^{i}$ (that allows us to write
$\partial_{[a}e_{b]}^{i}=-\epsilon^{i}_{\;\;jk}\Gamma_{[a}^{j}e_{b]}^{k}$).
With this in mind, and taking into account that (\ref{0015}) generates the
canonical transformations from $SO(3)$-ADM to the Ashtekar formalism we
expect that the action (\ref{2.1}) leads to $SO(3)$-ADM (as it turns out
to be the case). We follow now the usual procedure to get the Hamiltonian
formulation. The 3+1 decomposition gives
$$
\hspace{-14mm}
S={\displaystyle \int} dt{\displaystyle \int}_{\Sigma} d^{3}\!x \left\{
(2{\cal L}_{t} e_{a}^{i})
\tilde{\eta}^{abc}\nabla_{b}e_{ci}-2({\cal
L}_{t}A_{a}^{i})\tilde{\eta}^{abc}e_{bi}v_{c}+v_{0}
\tilde{\eta}^{abc}e_{a}^{i}F_{bci}-
\right.
$$
\begin{equation}
\hspace{1.2cm}
\left.-A_{0}^{i}\nabla_{a}\left[\tilde{\eta}^{abc}(
\epsilon_{i}^{\;\;jk}e_{bj}e_{ck}-2v_{b}e_{ci})\right]
+e_{0}^{i}\;\tilde{\eta}^{abc}
\left[\epsilon_{i}^{\;\;jk}F_{abj}e_{ck}-v_{a}F_{bci}\right]\right\}
\label{2.2}
\end{equation}
{}From (\ref{2.2}) we get the following primary constraints
\begin{equation}
\hspace{-2.3cm}
\begin{array}{l}
\tilde{\pi}_{i}=0 \\
\tilde{\sigma}_{i}=0 \\
\tilde{p}=0
\end{array}
\label{100a}
\end{equation}
\begin{equation}
\begin{array}{l}
\tilde{\pi}_{i}^{a}+2\tilde{\eta}^{abc}e_{bi}v_{c}=0 \\
\tilde{\sigma}_{i}^{a}-2\tilde{\eta}^{abc}\nabla_{b}e_{ci}=0\\
\tilde{p}^{a}=0
\end{array}
\label{2.3}
\end{equation}
We define now a total Hamiltonian $H_{T}$ by adding the primary
constraints (multiplied by Lagrange multipliers $u^{i}$, $u_{a}^{i}$,
$v^{i}$, $v_{a}^{i}$, $w$, and $w_{a}$) to the Hamiltonian derived from
(\ref{2.2})
$$
H_{T}={\displaystyle \int}_{\Sigma} d^{3}\!x \left\{
A_{0}^{i}\nabla_{a}\left[\tilde{\eta}^{abc}
(\epsilon_{i}^{\;\;jk}e_{bj}e_{ck}-2v_{b}e_{ci})\right]
-e_{0}^{i}\;\tilde{\eta}^{abc}
\left[\epsilon_{i}^{\;\;jk}F_{abj}e_{ck}-v_{a}F_{bci}\right]
\right.+
$$
\begin{equation} \hspace{11mm}
-v_{0}\tilde{\eta}^{abc}F_{abi}e_{c}^{i}+u^{i}\tilde{\pi}_{i}+u_{ai}
\left[\tilde{\pi}_{i}^{a}+2\tilde{\eta}^{abc}
e_{bi}v_{c}\right]+v^{i}\tilde{\sigma}_{i}+v_{a}^{i}\left[
\tilde{\sigma}_{i}^{a}-
2\tilde{\eta}^{abc}\nabla_{b}e_{ci}\right]+\hspace{5mm} \label{2.4}
\end{equation}
$$
\hspace{-10cm}\left.+w\tilde{p}+w_{a}\tilde{p}^{a}\right\}
$$
The conservation in time of the primary constraints under the evolution
given by $H_{T}$ gives the following secondary constraints
\begin{eqnarray}
& & \nabla_{a}\left[\tilde{\eta}^{abc}
(\epsilon_{i}^{\;\;jk}e_{bj}e_{ck}-2v_{b}e_{ci})\right]=0\nonumber\\
& & \tilde{\eta}^{abc}
\left[\epsilon_{i}^{\;\;jk}F_{abj}e_{ck}-v_{a}F_{bci}\right]=0\label{2.5}\\
& & \tilde{\eta}^{abc}F_{abi}e_{c}^{i}=0\nonumber
\end{eqnarray}
When added to the set of primary constraints they are second class. As
usual, it is possible to find linear combinations of the second class
constraints that are first class by solving some consistency equations for
the Lagrange multipliers introduced in $H_{T}$. For example, we can show
that each of the secondary constraints (\ref{2.5}) will give rise to a
first class constraint in the final formulation. In addition to these,
there is an additional first class constraint (responsible for generating
the $SO(3)$ factor in $SO(4)$ that is usually gauged away) given by
\begin{equation}
(v_{a}\delta_{ik}+\epsilon_{ijk}e_{a}^{j})\left[\tilde{\sigma}_{k}^{a}
-2\tilde{\eta}^{abc}\nabla_{b}e_{ck}\right]-e_{ck}\tilde{p}^{c}=0
\label{2.6}
\end{equation}
As commented above it is possible to gauge away $v_{a}$ and thus, remove
both $v_{a}$ and $\tilde{p}^{a}$ from the final canonical formulation
(this can be done also by imposing the gauge fixing condition $v_{a}=0$
and solving $\tilde{p}^{a}=0$). After doing this we are left with the
second class constraints
\begin{eqnarray}
& & \tilde{\pi}^{a}_{i}=0\label{2.7.1}\\
& & \tilde{\sigma}_{k}^{a}
-2\tilde{\eta}^{abc}\nabla_{b}e_{ck}=0\label{2.7.2}
\end{eqnarray}
--that must be solved-- and the constraints
\begin{eqnarray}
& &
\nabla_{a}\left[\tilde{\eta}^{abc}\epsilon_{ijk}e_{b}^{j}e_{c}^{k}
\right]=0\label{2301}\\
& & \tilde{\eta}^{abc}\epsilon_{ijk}F_{ab}^{j}e_{c}^{k}=0\label{2302}\\
& & \tilde{\eta}^{abc}F_{abi}e_{c}^{i}=0\label{2303}
\end{eqnarray}
Introducing the solution to (\ref{2.7.1}, \ref{2.7.2}) in (\ref{2301}-
\ref{2303}), they will become first class.  Equation (\ref{2.7.1})
suggests that the best thing to do, at this point, is to write the
connection $A_{a}^{i}$ in terms of $\tilde{\sigma}_{k}^{a}$ and $e_{ai}$
by using $\tilde{\sigma}_{k}^{a}-2\tilde{\eta}^{abc}\nabla_{b}e_{ck}=0$.
In this way we can rewrite all the constraints (\ref{2301}-\ref{2303}) in
terms of the canonically conjugate pair of variables
$\tilde{\sigma}_{k}^{a}$ and $e_{ai}$. Notice that the only place in which
the condition $\tilde{\pi}^{a}_{i}=0$ must be taken into account is in the
symplectic structure\footnote{$d\!l$ represents the generalized exterior
differential in the infinite-dimensional phase space spanned by
$A_{a}^{i}(x)$, $e_{a}^{i}(x)$, $\tilde{\pi}^{a}_{i}(x)$, and
$\tilde{\sigma}^{a}_{i}(x)$.}
\begin{equation}
\Omega=\int_{\Sigma} d^3\!x \left[d\!l\tilde{\pi}^{a}_{i}(x)\wedge
d\!lA_{a}^{i}(x)+d\!l\tilde{\sigma}^{a}_{i}(x)\wedge d\!le_{a}^{i}(x)\right]
\label{2.9}
\end{equation}
where it cancels the first term. This means that $\tilde{\sigma}_{k}^{a}$
and $e_{ai}$
are indeed a canonical pair of variables in the final phase space. The
solution to
$\tilde{\sigma}_{k}^{a}-2\tilde{\eta}^{abc}\nabla_{b}e_{ck}=0$ is
\begin{equation}
A_{a}^{i}=\Gamma_{a}^{i}+K_{a}^{i}
\label{2.10}
\end{equation}
where $\Gamma_{a}^{i}$ and $K_{a}^{i}$ are given by
\begin{equation}
\Gamma_{a}^{i}=-\frac{1}{2\tilde{e}}(e_{a}^{i}e_{b}^{j}-2e_{a}^{j}e_{b}^{i})
\tilde{\eta}^{bcd}\partial_{c}e_{dj}
\label{2.11}
\end{equation}
\begin{equation}
\hspace{-8mm}K_{a}^{i}=\frac{1}{4\tilde{e}}(e_{a}^{i}e_{b}^{j}-
2e_{a}^{j}e_{b}^{i})
\tilde{\sigma}^{b}_{j}
\label{2.12}
\end{equation}
($\tilde{e}\equiv \frac{1}{6}
\tilde{\eta}^{abc}\epsilon_{ijk}e_{a}^{i}e_{b}^{j}e_{c}^{k}$ is
the determinant of the triad). It is straightforward to show that the
previous $\Gamma_{a}^{i}$ is compatible with $e_{a}^{i}$ (i.e. ${\cal
D}_{a}e_{b}^{i}\equiv\partial_{a}e_{b}^{i}-\Gamma_{ab}^{c}e_{c}^{i}+
\epsilon^{i}_{\;\;j
k}\Gamma_{a}^{j}e_{b}^{k}=0$ where $\Gamma_{ab}^{c}$ are the Christoffel
symbols built with the three dimensional metric $q_{ab}\equiv
e_{a}^{i}e_{bi}$).

Equation in (\ref{2301}) gives immediately (just substituting
$2\tilde{\eta}^{abc}\nabla_{b}e_{ck}=\tilde{\sigma}_{k}^{a}$)
\begin{equation}
\epsilon_{ijk}e_{a}^{j}
\tilde{\sigma}^{ak}=0
\label{2.13}
\end{equation}
This is the generator of $SO(3)$ rotations. Differentiating now in equation
(\ref{2.7.2}) we get
$\nabla_{a}\tilde{\sigma}^{a}_{i}=2\tilde{\eta}^{abc}\nabla_{a}\nabla_{b}e_{ci}
=\tilde{\eta}^{abc}\epsilon_{ijk}F_{ab}^{j}e_{c}^{k}=0$ where we have made use
of
(\ref{2302}). In order to eliminate the $A_{a}^{i}$ from
$\nabla_{a}\tilde{\sigma}^{a}_{i}$ we add and subtract
$\epsilon_{i}^{\;\;jk}\Gamma_{aj}\tilde{\sigma}^{a}_{k}$  to get ${\cal
D}_{a}\tilde{\sigma}^{a}_{i}=-\epsilon_{i}^{\;\;jk}K_{a}^{j}
\tilde{\sigma}^{a}_{k}$.
It is straightforward to show that the right hand side of this last
expression is
zero by using the definition of $K_{a}^{i}$ and the ``Gauss law"
(\ref{2.13}). We have
then
\begin{equation}
{\cal D}_{a}\tilde{\sigma}^{a}_{i}=0
\label{2.14}
\end{equation}
Finally, the scalar constraint is obtained by introducing $F_{ab}^{i}=
R_{ab}^{i}+2{\cal
D}_{[a}K_{b]}^{i}+\epsilon^{i}_{\;\;jk}K_{a}^{j}K_{b}^{k}$ (where
$R_{ab}^{i}\equiv2\partial_{[a}\Gamma_{b]}^{i}+\epsilon^{i}_{\;\;jk}
\Gamma_{a}^{j}
\Gamma_{b}^{k}$ is the curvature of $\Gamma_{a}^{i}$) in
(\ref{2303}) and using the Gauss law (\ref{2.13}). The final result is
\begin{equation}
\tilde{\eta}^{abc}R_{ab}^{i}e_{ci}+\frac{1}{8\tilde{e}}
\left[e_{a}^{i}e_{b}^{j}-2e_{a}
^{j}e_{b}^{i}\right]\tilde{\sigma}^{a}_{i}\tilde{\sigma}^{b}_{j}=0
\label{2.15}
\end{equation}

In order to connect this to the usual ADM and $SO(3)$-ADM formalisms we first
write
\begin{equation}
\begin{array}{l}
q_{ab}=e_{a}^{i}e_{bi}\\
K_{ab}={\displaystyle
\frac{1}{4\tilde{e}}}\left[2q_{c(a}e_{b)}^{i}\tilde{\sigma}^{c}_{i}-
q_{ab}e_{c}^{i}\tilde{\sigma}^{c}_{i}\right]
\end{array}
\label{2.16}
\end{equation}
Taking into account that $\tilde{p}^{ab}=\sqrt{\tilde{\!\tilde{q}}}
(K^{ab}-K\;q^{ab})$
($\;\tilde{\!\tilde{q}}$ is the determinant of the 3-metric
$q_{ab}\equiv e_{a}^{i}e_{bi}$)
we find that (\ref{2.16}) implies\footnote{$e^{a}_{i}$ is the inverse
of $e_{ai}$.}
\begin{equation}
\tilde{p}^{ab}=\frac{1}{2}e^{(a}_{i}\tilde{\sigma}^{b)i}
\label{2.17}
\end{equation}
These expressions allow us to immediately check that $q_{ab}$ and
$\tilde{p}^{ab}$ are a pair of canonically conjugate variables. By using
the ``Gauss law" (\ref{2.13}) we can remove the symmetrizations in
(\ref{2.17}) and write
$\tilde{p}^{ab}=\frac{1}{2}e^{b}_{i}\tilde{\sigma}^{ai}$. With this last
expression it is straightforward to show that the constraint (\ref{2.14})
gives ${\cal D}_{a}\tilde{p}^{ab}=0$ (i.e. the familiar vector constraint
in the ADM formalism). Because there is no internal symmetry in the usual
ADM formalism the only thing we are left to compute is the scalar
constraint. From (\ref{2.15}) and using the fact that
$R=-\epsilon^{ijk}R_{abi}e^{a}_{j}e^{b}_{k}=-\frac{1}{\tilde{e}}
\tilde{\eta}^{abc}
R_{ab}^{i}e_{ci}$ and $\tilde{\sigma}_{i}^{a}=2\tilde{p}^{ab}e_{bi}$
--modulo (\ref{2.13})-- we get
\begin{equation}
-\sqrt{\tilde{\!\tilde{q}}}R+\frac{1}{\sqrt{\tilde{\!\tilde{q}}}}\left(
\frac{1}{2}
\tilde{p}^2-\tilde{p}^{ab}\tilde{p}_{ab}\right)=0
\label{2.18}
\end{equation}
The relative signs between the potential and kinetic terms in the previous
expression
correspond to Euclidean signature if we take real fields.

In order to see how our result gives the $SO(3)$-ADM formalism of ref.
\cite{eta} we
write\footnote{$\tilde{\tilde {\pi}}\equiv \det \tilde{\pi}^{a}_{i}$.}
\begin{eqnarray}
& & \tilde{\pi}^{a}_{i}=\frac{\mu}{2}\tilde{\eta}^{abc}\epsilon_{ijk}
e_{b}^{j}e_{c}^{k}
\label{301}\\
& &
K_{a}^{i}=\frac{1}{2\mu\tilde{e}}\left(e_{a}^{i}e_{b}^{j}-2e_{a}^{j}
e_{b}^{i}\right)
\tilde{\sigma}^{b}_{j}\label{302}
\end{eqnarray}
and their inverses
\begin{eqnarray}
& & e_{ai}=\frac{1}{2
\sqrt{\mu\tilde{\tilde{\pi}}}}\;\;\eut_{abc}\epsilon^{ijk}\tilde{\pi}^{b}_{j}
\tilde{\pi}^{c}_{k} \label{303}\\
& &\tilde{\sigma}^{a}_{i}=2\sqrt{\frac{\mu}{\tilde{\tilde{\pi}}}}
\tilde{\pi}^{[a}_{i}
\tilde{\pi}^{c]}_{k}K_{c}^{k}\label{304}
\end{eqnarray}
It is straightforward to check that these equations define a canonical
transformation
for every value of the arbitrary constant $\mu$; the relevant Poisson
 bracket is
\begin{equation}
\left\{\tilde{\pi}^{a}_{i}(x),
K_{b}^{j}(y)\right\}=\delta_{b}^{a}\delta_{i}^{j}\delta^{3}(x, y)
\label{2001}
\end{equation}
In the following I will use the inverse of $\frac{\tilde{\pi}^{a}_{i}}
{\tilde{\tilde{\pi}}}$ that I will denote $E_{a}^{i}$. Substituting (\ref{303},
\ref{304}) in the constraints (\ref{2.13}-\ref{2.14})
we easily get the ``Gauss law" and the vector constraint
\begin{eqnarray}
& & \epsilon_{ijk}K_{a}^{j}\tilde{\pi}^{ak}=0\nonumber\\
& & {\cal D}_{a}\left[\tilde{\pi}^{a}_{k}K_{b}^{k}-
\delta_{b}^{a}\tilde{\pi}^{c}_{k}
K_{c}^{k}\right]=0\label{305}
\end{eqnarray}
In order to get the Hamiltonian constraint we need
\begin{equation}
\tilde{\eta}^{abc}R_{ab}^{i}(e)e_{ci}=\frac{1}{2\sqrt{\mu
\tilde{\tilde{\pi}}}}\eta^{abc}
R_{ab}^{i}(e)\;\eut_{cde}\epsilon_{i}^{\;\;jk}\tilde{\pi}^{d}_{j}
\tilde{\pi}^{f}_{k}=
\frac{1}{\sqrt{\mu}}R_{ab}^{i}(E)\tilde{\eta}^{abc}E_{ci}=-
\sqrt{\frac{\tilde{\!\tilde{q}}}{\mu}}R
\end{equation}
where $\tilde{\!\tilde{q}}^{ab}=\tilde{\pi}^{a}_{i}\tilde{\pi}^{bi}$ and
we have used the
fact that $R_{ab}^{i}(e)=R_{ab}^{i}(E)$. We also need
\begin{equation}
\frac{1}{8\tilde{e}}\left[e_{a}^{i}e_{b}^{j}-2e_{a}^{j}e_{b}^{i}\right]
\tilde{\sigma}^{a}_{i}\tilde{\sigma}^{b}_{j}=\frac{\mu^{3/2}}
{4\tilde{\tilde{\pi}}}\left[(\tilde{\pi}^{a}_{i}K_{a}^{i})^{2}-
(\tilde{\pi}^{a}_{i}K_{a}^{j})(\tilde{\pi}^{b}_{j}K_{b}^{i})\right]
\end{equation}
to finally get the following Hamiltonian constraint
\begin{equation}
\sqrt{\tilde{\!\tilde{q}}}R+\frac{\mu^{2}}{2\sqrt{\tilde{\!\tilde{q}}}}
\tilde{\pi}^{[b}_{i}
\tilde{\pi}^{a]}_{j}K_{a}^{i}K_{b}^{j}=0
\label{2002}
\end{equation}
By choosing $\mu=2$ we find the result of \cite{eta}.

The action introduced in this section has some nice characteristics. It
shows, for example, that it is possible to get an action for the
``geometrodynamical" Husain-Kucha\v{r} model simply by removing the term with
$v_{a}$. It can also be used to discuss the issue of reality conditions
and to get one of the
real Ashtekar formulations for Lorentzian gravity presented in \cite{fer1}.
This will be the scope of the next two sections.

\section{Reality conditions}

In this section I will show how the action (\ref{2.1}) can be used to
discuss the reality conditions of the theory and the signature of the
space-time. The starting point is realizing that since we are working with
complex fields, multiplying (\ref{2.1}) by a purely imaginary factor (say
i) cannot have any effect on the theory (because the field equations
will remain unchanged). However, same produces some changes in the
Hamiltonian formulation. Following the derivation presented in the previous
section we find now that the primary constraints are
\begin{equation}
\hspace{-2.3cm}
\begin{array}{l}
\tilde{\pi}_{i}=0 \\
\tilde{\sigma}_{i}=0 \\
\tilde{p}=0
\end{array}
\label{10a}
\end{equation}
\begin{equation}
\begin{array}{l}
\tilde{\pi}_{i}^{a}+2i\tilde{\eta}^{abc}e_{bi}v_{c}=0 \\
\tilde{\sigma}_{i}^{a}-2i\tilde{\eta}^{abc}\nabla_{b}e_{ci}=0\\
\tilde{p}^{a}=0
\end{array}
\label{2.300}
\end{equation}
whereas the secondary constraints are still given by (\ref{2.5}). Fixing
the gauge
$v_{a}=0$ we have
\begin{eqnarray}
& & \tilde{\pi}^{a}_{i}=0\label{2.19.1}\\
& & \tilde{\sigma}_{k}^{a}
-2i\tilde{\eta}^{abc}\nabla_{b}e_{ck}=0\label{2.19.2}
\end{eqnarray}
together with (\ref{2301}-\ref{2303}). As we did before we must solve
the second class
constraints (\ref{2.19.1})(\ref{2.19.2}). The connection $A_{a}^{i}$ is
now given by
\begin{equation}
A_{a}^{i}=\Gamma_{a}^{i}-\frac{i}{4\tilde{e}}(e_{a}^{i}e_{b}^{j}-2e_{a}^{j}
e_{b}^{i})
\tilde{\sigma}^{b}_{j}
\label{2.20}
\end{equation}
Although the solution for $A_{a}^{i}$ has some imaginary factors,
it is clear that nothing will change in the symplectic
structure because we have $\tilde{\pi}^{a}_{i}=0$. In order to get the
expressions for
the constraints we can simply make the change
$\tilde{\sigma}^{a}_{i}\rightarrow -i\tilde{\sigma}^{a}_{i}$ in
(\ref{2.13}-\ref{2.15}). The Gauss law and the vector constraint will be
unaffected because they are linear and homogeneous in
$\tilde{\sigma}^{a}_{i}$. There is, however, a change in the Hamiltonian
constraint precisely in the relative sign between the kinetic and
potential terms (due to the fact that it is not a homogeneous polynomial
in $\tilde{\sigma}^{a}_{i}$). Actually we get
\begin{equation}
\tilde{\eta}^{abc}R_{ab}^{i}e_{ci}-\frac{1}{8\tilde{e}}\left[e_{a}^{i}
e_{b}^{j}-2e_{a}
^{j}e_{b}^{i}\right]\tilde{\sigma}^{a}_{i}\tilde{\sigma}^{b}_{j}=0
\label{2.21}
\end{equation}
If the fields in (\ref{2.21}) are taken to be real, the relative sign
between the two terms in this Hamiltonian constraint corresponds to
Lorentzian general relativity. Notice, however, that, as long as we remain
within the realm of the complex theory the signature of the space-time is
not defined. It is only when we add the reality conditions that we pick a
signature or the other.  The conclusion that we draw from this fact
is that, when we use the modified self-dual action, the signature
of the space-time in the real formulation is controlled both by the form of the
Hamiltonian constraint and the reality conditions.
If we start form (\ref{2.15}) the reality conditions \begin{equation}
\begin{array}{l}
e_{a}^{i}\;\;\;\;{\rm real}\\
\tilde{\sigma}^{a}_{i}\;\;\;\;{\rm real}
\label{2.22}
\end{array}
\end{equation}
and
\begin{equation}
\begin{array}{l}
e_{a}^{i}\;\;\;\;{\rm real}\\
\tilde{\sigma}^{a}_{i}\;\;\;\;{\rm purely \;\;imaginary}
\label{2.23}
\end{array}
\end{equation}
lead to Euclidean and Lorentzian general relativity respectively, whereas
if we use (\ref{2.21}) then the reality conditions (\ref{2.22}) give
Lorentzian signature and (\ref{2.23}) Euclidean signature.

We could do the same thing starting from the self-dual action written
in the form
(\ref{7}), we would find, then, that the  primary constraints (\ref{10b})
 become
\begin{equation}
\begin{array}{l}
\tilde{\pi}_{i}^{a}-i\tilde{\eta}^{abc}(2v_{b}e_{ci}-
\epsilon_{i}^{\;\;jk}e_{bj}e_{ck})
=0 \\
\tilde{\sigma}_{i}^{a}=0 \\
\tilde{p}^{a}=0
\end{array}
\label{2.24}
\end{equation}
whereas the rest of the primary and secondary constraints are left
unchanged\footnote{as above, we will fix the gauge $v_{a}=0$.}. As we have
$\tilde{\sigma}^{a}_{i}=0$ nothing will change in the symplectic structure
after writing $e_{a}^{i}$ in terms of $\tilde{\pi}^{a}_{i}$. The new
constraints can be obtained now by making the transformation
$\tilde{\pi}^{a}_{i}\rightarrow -i\tilde{\pi}^{a}_{i}$ in (\ref{0014}). As
they are homogeneous polynomials in the momenta we get exactly the result
that we found in section II. We have then an interesting situation. With
the modified self-dual action we have two alternative expressions for the
constraints and two different sets of reality conditions that we can use
to control the signature of the space-time in the final real formulation.
If we use the self-dual action, however, it seems that we can only control
the signature by using reality conditions. It turns out that this is not
strictly true. In fact, as it happens in the previous example, there are
actually several possible ways to write the Hamiltonian constraint in
terms of Ashtekar variables with trivial reality conditions \cite{fer1},
some of them for Lorentzian signatures and some for Euclidean signatures.
This is the subject of the next section.

\section{A real Lorentzian formulation}

I show in this section that it is possible to use the constraint analysis
of the modified self-dual action to obtain a real formulation with
Ashtekar variables for Lorentzian signature space-times.

As already pointed out in the introduction there are two key issues that
lead to the success of the Ashtekar approach to classical and quantum
gravity. One of them is the geometrical nature of the new variables, the
other the simple polynomiality of the constraints (specially the scalar
constraint). Many of the insights about quantum gravity gained with the
new formalism have to do with the use of loop variables. Among them the
introduction of the area and volume observables \cite{SSS}, \cite{SRA} and
the construction of weave states are very interesting because they give
physical predictions about the structure of space-time at the Planck
scale. Unfortunately, the implementation of the reality conditions in the
loop variables framework is rather difficult in the absence of an explicit
real formulation and then the results obtained had to be accepted only
modulo the reality conditions.  For this reason it is very desirable to
formulate the theory in terms of a real Ashtekar connections and triads.
If this can be achieved then it is possible to argue that all the results
obtained within the loop variables framework which do not require the use
of the scalar constraint are true without having to worry about the issue
of reality conditions. This applies, for example to the area and volume
observables studied mentioned before.

In retrospective, getting a real formulation for Lorentzian signature
space-times in terms of Ashtekar variables is very easy once we accept to
live with a more complicated Hamiltonian constraint. In fact, one can just
take the scalar constraint (\ref{2.21}) (corresponding to Lorentzian
signature if written in terms of real fields) and substitute
$\tilde{\sigma}_{i}^{a}$ and $e_{a}^{i}$ for their values in terms of
$A_{a}^{i}$ and $\tilde{\pi}^{a}_{i}$ given by (\ref{303},\ref{304}),
namely
\begin{eqnarray}
& & e_{ai}=\frac{1}{2\sqrt{2\tilde{\tilde{\pi}}}}
\;\;\eut_{abc}\epsilon^{ijk}\tilde{\pi}^{b}_{j}
\tilde{\pi}^{c}_{k} \label{2.27}\\
& &\tilde{\sigma}^{a}_{i}=2\sqrt{\frac{2}{\tilde{\tilde{\pi}}}}
\tilde{\pi}^{[c}_{i}
\tilde{\pi}^{a]}_{k}(A_{c}^{k}-\Gamma_{c}^{k})\label{2.28}
\end{eqnarray}
The Gauss law and the vector constraints are the usual ones whereas
the Hamiltonian
constraint becomes
\begin{eqnarray}
\hspace{-1.5cm}\epsilon^{ijk}\tilde{\pi}^{a}_{i}\tilde{\pi}^{b}_{j}F_{abk}-
\tilde{\eta}^{a_{1}a_{2}a_{3}}\tilde{\eta}^{b_{1}b_{2}b_{3}}\left[
(e_{a_{1}}^{i}
\nabla_{a_{2}}e_{a_{3}i}) (e_{b_{1}}^{j}\nabla_{b_{2}}e_{b_{3}j})\right. & &
\label{2.29} \\
& \left.\hspace{-2.5cm}-2(e_{a_{1}}^{j}\nabla_{a_{2}}e_{a_{3}i})
(e_{b_{1}}^{i}\nabla_{b_{2}}e_{b_{3}j})\right]=0 &
\nonumber
\end{eqnarray}
where $e_{ai}$ must be written in terms of $\tilde{\pi}^{a}_{i}$ from
(\ref{2.27}). As we can see, in addition to the usual term there is
another involving covariant derivatives of $\tilde{\pi}^{a}_{i}$. The same
result is obtained if we do not use any imaginary unit in the canonical
transformations that connects the $SO(3)$-ADM and Ashtekar formulations
and we start from the Lorentzian ADM scalar constraint. An interesting
fact (and useful when checking that we get the right constraint algebra)
is that $e_{ai}[\tilde{\pi}]$ and $-2\tilde
{\eta}^{abc}\nabla_{b}e_{ci}[\tilde{\pi}]$ are canonically conjugate
objects. This suggests the possibility of extending the usual loop
variables with the addition of objects depending on
$\tilde{\eta}^{abc}\nabla_{b}e_{ci}[\tilde{\pi}]$. This may be useful in
order to write the new Hamiltonian constraint in terms of them.

As we discussed before we can use both the form of the constraints and the
reality conditions to control the signature of the space-time. This means
that both the familiar Hamiltonian constraint in the Ashtekar formulation
and (\ref{2.29}) can be used to describe any space-time signature by
choosing appropriate reality conditions. Of course, if these two forms of
the Hamiltonian constraint are used for Euclidean and Lorentzian
signatures respectively, the reality conditions will be very simple (just
the condition that the fields be real). If, instead, the usual Hamiltonian
constraint is used for Lorentzian signatures or the new one for Euclidean
ones then the reality conditions will be more complicated.

\section{Conclusions and outlook}

By using a modified form of the self-dual action that leads to the
$SO(3)$-ADM formalism without the appearance of difficult second class
constraints we have studied the reality conditions of the theory and
obtained a real formulation in terms of Ashtekar variables for Lorentzian
signature space-times. We have been able to show that, both in the ADM and
Ashtekar phase spaces, it is possible to find different forms for the
Hamiltonian constraints for complexified general relativity. In order to
pass to a real formulation we need to impose reality conditions that can
be chosen to pick the desired space-time signature. In a sense, it is no
longer necessary to talk about reality conditions because we can impose
the trivial ones (real fields) and control the signature of the space time
by choosing appropriate Hamiltonian constraints.

The fact that a real formulation in the Ashtekar phase space is available
means that all the results obtained by using loop variables that are
independent of the detailed form of the Hamiltonian constraint are true
without having to worry about reality conditions. On the other hand there
a price to be paid; namely, that the Hamiltonian constraint is no longer a
simple quadratic expression in both the densitized triad and the Ashtekar
connection. This makes it more difficult to discuss all those issues that
depend critically on having the theory formulated in terms of simple
constraints; in particular solving the constraints will be more difficult
now.

In this respect one can honestly say that the structure of the Hamiltonian
constraint presented above (or the alternative forms discussed in
\cite{fer1}) is, at least, as complicated as the one of the familiar ADM
constraint. In spite of that, some interesting and basic features of the
Ashtekar formulation are retained. The phase space still corresponds to
that of a Yang-Mills theory, so we can continue to use loop variables in
the passage to the quantum theory. The ``problem" of reality conditions
has now been transformed into that of writing the new (and complicated)
Hamiltonian constraint in terms of loop variables and solving the quantum
version of the constraints acting on the wave functional. The final
success of this approach will depend on the possibility of achieving this
goal.

One interesting point of discussion suggested by the results presented in
the paper has to do with the obvious asymmetry between the formulations of
gravity in a real Ashtekar phase space for Lorentzian and Euclidean
signatures. In the geometrodynamical approach, there is little difference,
both in the Lagrangian and Hamiltonian formulations, between them (at
least at the superficial level of the complication of the expressions
involved). In fact it all boils down to the relative signs between the
potential and kinetic terms in the scalar constraint.  In our case,
however, the formulations that we get are, indeed, rather different.

The existence of the formulation presented in this paper also suggests
that the origin of the signature at the Lagrangian level is also rather
obscure. From the (real) self-dual-action it seems quite natural to
associate, for example, the euclidean signature with the fact that the
gauge group is $SO(4)$ and the metric is $e_{aI}e^{I}_{b}$. However, the
observation that one of the SO(3) factors ``disappears" from the theory
may be telling us that, perhaps, it is not necessary to start with an
$SO(4)$ internal symmetry. In fact the Capovilla-Dell-Jacobson \cite{CDJ}
Lagrangian leads to the same formulation using only $SO(3)$ as the
internal symmetry.

Although we still do not have a four dimensional Lagrangian formulation of
the theory, the marked asymmetry between the real Hamiltonian formulations
for different space-time signature strongly suggests that it would differ
very much from the usual self-dual action (a fact also supported by the
lack of success of all the attempts to get Lorentzian general relativity
by introducing simple modifications in the known actions). This could have
intriguing consequences in a perturbative setting because the UV behavior
(controlled to a great extent by the functional form of the Lagrangian) of
the Euclidean and the Lorentzian theories could be very different. It is
worthwhile to remember at this point that the Einstein-Hilbert action and
the so called higher derivative theories, that differ in some terms
quadratic in the curvatures, have very different UV behaviors. The first
one is non-renormalizable whereas the second one is renormalizable but
non-unitary. In our opinion this is an issue that deserves further
investigation.

{\bf Acknowledgements} I wish to thank A. Ashtekar, Ingemar Bengtsson,
Peter Peld\'an, Lee Smolin and Chopin Soo for several interesting
discussions and remarks. I am also grateful to the Consejo Superior de
Investigaciones Cient\'{\i}ficas (Spanish Research Council )for providing
financial support.

\end{document}